\documentclass[12pt]{article}

\begin{document}

\sloppy
 \begin{flushright}{SIT-HEP/TM-13}
 \end{flushright}
\vskip 1.5 truecm
\centerline{\large{\bf Thermal hybrid inflation in brane world}}
\vskip .75 truecm
\centerline{\bf Tomohiro Matsuda
\footnote{matsuda@sit.ac.jp}}
\vskip .4 truecm
\centerline {\it Laboratory of Physics, Saitama Institute of
 Technology,}
\centerline {\it Fusaiji, Okabe-machi, Saitama 369-0293, 
Japan}
\vskip 1. truecm
\makeatletter
\@addtoreset{equation}{section}
\def\theequation{\thesection.\arabic{equation}}
\makeatother
\vskip 1. truecm

\begin{abstract}
\hspace*{\parindent}
In conventional scenario of thermal inflation, the requirement that the
 reheating temperature must be larger than the temperature of the
 nucleosynthesis puts a lower bound on the mass of the inflaton field.  
At the same time, the mass of the inflaton field
and the height of the potential during thermal inflation are intimately
 related. 
With these conditions, the conventional models for thermal inflation are
 quite restricted. 
Naively, one can expect that the above constraints may be removed if
 thermal inflation is realized within the setups for hybrid 
inflation.
In this paper we show why it is difficult to construct hybrid models
for thermal inflation within conventional supergravity, and then show
a successful example in models for the braneworld.
Our mechanism is based on the idea of non-tachyonic brane inflation.
\end{abstract}

\newpage
\section{Introduction}
\hspace*{\parindent}
In conventional models for thermal inflation, inflation starts from the
state where the symmetry is restored by the thermal effect.
One may regard it as a modification of new inflation, where inflation
starts at the top of the potential.
On the other hand, hybrid inflation starts with large field initial
condition.
During hybrid inflation, the inflaton field stabilizes the trigger field
at the false vacuum.
In this respect, hybrid inflaton is similar to chaotic inflation.
Taking these things into consideration, it seems rather difficult to
construct thermal hybrid inflation without adding unnatural components 
by hand.
What we want to show in this paper is that the collaboration between
these different kinds of inflation can be realized in the natural
settings of brane inflation. 
\footnote{In general, inflation with low fundamental scale is very
difficult\cite{low_inflation}. Baryogenesis in models with low
fundamental scale is discussed in ref.\cite{low_baryo}.}

\section{Thermal hybrid inflation}
\hspace*{\parindent}
In this section we first consider an old idea of inverted hybrid
inflation\cite{inverted} where the hybrid potential is used while the
initial field expectation value is not large.

\underline{Inverted hybrid inflation}

In the scenario for inverted inflation, the inflaton field 
slowly rolls away from the origin, and finally   
the trigger field terminates inflation.
Thus at first sight, it seems possible to construct thermal hybrid 
inflation, if it is realized in the setups of the inverted scenario.
Here we examine the above simplest idea and show that it is indeed quite
difficult to be realized without fine-tunings.

Perhaps the simplest way to explain the idea of inverted hybrid
inflation is to consider the potential 
\begin{equation}
V(\phi, \sigma)=\frac{1}{4}\lambda_{\sigma}\left(\sigma^2 + M^2 \right)^2 
-\frac{1}{2}\lambda_{int} \phi^2 \sigma^2
+\frac{1}{4}\lambda_{\phi}\left(\phi^2-M'^2\right)^2
\end{equation}
where $\phi$ is the inflation field and $\sigma$ is the trigger field.
Inverted hybrid inflation starts when the inflaton field ($\phi$) stays at
the top of the potential ($\phi\simeq 0$), where the trigger 
field is stabilized at $\sigma =0$.
Here we examine if the thermal initial condition works well for this model.
We assume that, because of the thermal effect, $\phi$ is held at the
origin during thermal inflation. 
Since the inflaton field ($\phi$) stays at $\phi=0$, the trigger field
($\sigma$) is also held at a false vacuum $\sigma=0$ until
$\phi$ reaches the critical value $\phi_c$.
Then the effective mass squared for the $\sigma$ field becomes negative,
allowing $\sigma$ to roll down to its true vacuum.
In this simplest model, however, the dimensionless coupling constant
$\lambda_\phi$ must be fine-tuned so that thermal inflation is allowed, 
which is the same situation as the original model for thermal
inflation\cite{thermal}.
Naively, one may consider a flat potential for $\phi$
\begin{equation}
\label{flat}
V(\phi, \sigma)=\frac{1}{4}\lambda_{\sigma}\left(\sigma^2 + M^2 \right)^2 
-\frac{1}{2}\lambda_{int} \phi^2 \sigma^2
-\frac{1}{2}m_{\phi}^2\phi^2 + A_n \frac{\phi^{n+4}}{M^n},
\end{equation}
assuming that the potential for the inflaton $\phi$ is flat in the
supersymmetric limit and is
destabilized by the soft supersymmetry breaking term, which is denoted by
$-\frac{1}{2}m_{\phi}^2\phi^2$.
In this case, the flatness of the potential seems to be ensured by
supersymmetry.
However, as is already discussed in ref.\cite{inverted_critical},
it is hard to obtain a flat potential while making the required
coupling large enough to destabilize the trigger field at the end of
inflation.
Moreover, if one wants to construct ``strong'' thermal
inflation with the number of e-foldings $N_e >35$, 
one should consider another constraint for $\lambda_{\sigma}M^4$.
During thermal inflation, the inflaton field is stabilized at the
symmetric point by the thermal effect.
Thermal inflation starts at $T_{in} \simeq
\left(\frac{1}{4}\lambda_\sigma M^4 \right)^{1/4}$ and ends at
$T_c \simeq m_{\phi}$.
The expansion during this period is $N_e \simeq 
ln\left(\frac{T_{in}}{T_c}\right)$.
In order to obtain large number of e-foldings, the vacuum energy during 
thermal inflation must be as large as $V_0^{1/4}\simeq m_\phi e^{N_e}$.
Within the setups of the conventional models for supergravity,
it seems very hard to satisfy any of the above constraints without adding 
unnatural extra components.

Thus we conclude that the model is not suitable for our purposes.
We will show that these problems are solved in the setups of brane
inflation.

\underline{Non-tachyonic brane inflation}

Here we consider non-tachyonic brane inflation in ref.\cite{matsuda_nontach2},
and examine whether one can construct a model that is suitable for our
purposes.
The model should be similar to the model of inverted hybrid inflation 
that we have discussed above.
For example, we consider a potential of the form
\begin{equation}
\label{simple}
V(\phi, \sigma)=\frac{1}{4}\lambda_{\sigma}\left(\sigma^2 + M^2 \right)^2 
-\frac{1}{2}\lambda_{int} e^{-(M_0 r)^2}\phi^2 \sigma^2
+\frac{1}{4}\lambda_{\phi}\left(\phi^2-M'^2\right)^2.
\end{equation}
Here we have considered two branes at a distance, which we denote by 
1 and 2. 
The field $\sigma$ and the field $\phi$ are localized on brane 1 and
brane 2, respectively. 
\footnote{
Here we consider the case in which fields are localized on the branes.
The forms of their wave-functions will be gaussian that decays
exponentially in the transverse directions.
For example, we consider the form;
\begin{equation}
\phi(r)\sim e^{(M_0 r)^2},
\end{equation}
where $r$ denotes the transverse distance from the center of the brane, 
and $M_0^{-1}$ is the width of the
wave-function.
When branes are separated, the couplings between localized modes from
different branes will decay exponentially.
On the other hand, when branes are on top of each other, the couplings 
are restored.
The effect of the localized wave-functions is included in the second 
term in eq.(\ref{simple}).}
The interaction term is accompanied by an exponential factor, because 
the field $\phi$ and the field $\sigma$ are localized on
different branes at a distance $r$.
On can easily find that the above effective four-dimensional potential
is a simple modification of the potential for inverted hybrid inflation.
In this model, the inflaton field is the moduli for the brane distance
$r$, which we denote by $\psi=M_*^2 r$.
The most obvious difference is that the field $\phi$ is not required to
be placed at the unstable point.
Inflation starts because the exponential factor is nearly zero at the
beginning of inflation.
If $\lambda_{int} M'^2 > \lambda_{\sigma}M^2$,
the field $\sigma$ is destabilized at the end of inflation, when two
branes come close.

It is possible to express the above idea (separation between branes
drives hybrid inflation) in the explicit supersymmetric
form\cite{matsuda_nontach2}.
The simplest example is the F-term brane inflation.
On one brane, we assume a localized superfield $\Sigma$ and its
superpotential of the form
\begin{equation}
W_1=\Sigma M^2.
\end{equation}
On the other brane, we assume that another field $\Phi$ is localized
and its superpotential is $W_2=0$.
When two branes come close, we assume another
superpotential of the form
\begin{equation}
W_{int}=\Sigma(M^2- \lambda_{i}\Phi^2 e^{-(M_0 r)^2}),
\end{equation}
where $r$ denotes the distance between branes.
One can explain the forms of the superpotential by imposing R-symmetry.
In eq.(\ref{simple}), the vacuum energy during inflation is positive
because $\sigma$ is located at the false vacuum ($\sigma=0$).
In the present case, the vacuum energy during inflation is due to the
supersymmetry breaking induced by $\Sigma$.
In both cases, the interaction is restored at the end of inflation.
In eq.(\ref{simple}), the potential for $\sigma$ is destabilized by
the interaction.
In the present case, supersymmetry restoration is induced by the
interaction term.

For an another example, we may consider a localized Fayet-Iliopoulos
D-term on a brane at $\vec{r}=0$,
\begin{equation}
\label{FI}
\xi D \delta(\vec{r})
\end{equation}
where $D$ is an auxiliary field of the vector superfield.
We consider an additional abelian gauge group $U(1)_X$ in the bulk,
while the Fayet-Iliopoulos term for $U(1)_X$ is localized on a brane. 
We also include the field $\phi_X$ that has $U(1)_X$ charge and
localized on the other brane at $\vec{r}=\vec{r_1}$.
Then the D-flat condition is
\begin{equation}
\xi \delta(\vec{r}) 
+c_X (|\phi^+_X|^2-|\phi^-_X|^2)\delta(\vec{r}-\vec{r_1})=0,
\end{equation}
where $c_X$ is a constant, and the $\delta$-functions are
gaussian located on each branes.
When two branes are placed at a distance ($|\vec{r_1}| >>M_*^{-1}$),
the Fayet-Iliopoulos term breaks supersymmetry on the brane,
and inflation starts.
In this case, as in the conventional models for brane inflation, 
the inflaton field is the moduli that parametrizes the
brane distance.
The moduli is denoted by $\psi=M_*^2 r_1$, 
where $M_*$ denotes the fundamental scale of the model. 

The issue of the potential for the brane distance is discussed by many authors 
for many models.
Of course, there are many contributions at the same time, which will take 
different forms.
For the typical example,
\begin{itemize}
\item The simple $m_{3/2}^2$ correction from the effective low energy
      discription of supergravity.
\item The loop corrections to the K\"ahler metric that comes from the 
particles of the mass $\sim M r^2$\cite{thermal_brane}.
\item Since the cancellation between the graviton-dilaton attraction and 
the RR repulsion fails when supersymmetry is broken, the Van der Waals
      forces between branes will appear\cite{thermal_brane}.
\end{itemize}

The trigger field is the localized field $\phi_X$, which develops vacuum
expansion value to compensate the D-term (\ref{FI}) when two branes come
close.

The most significant difference from the original model for inverted
hybrid inflation is that the interaction that destabilizes the trigger
field is accompanied by an exponential factor $e^{-(M^{-1}_{*} \psi)^2}$.
The destabilization in the potential (\ref{simple}) is not due to the
variation of the field $\phi$, 
but due to the variation of the effective interaction constant
$\lambda_{int}e^{-(M^{-1}_{*} \psi)^2}$.

In this case, however, the initial expectation value for the inflaton is
large.  
In this respect, to collaborate with thermal inflation, we should modify
the initial 
condition for the above extended model for hybrid inflation.
To find the solution for the problem, we must first review the original
idea for thermal brane inflation that was advocated in
ref.\cite{thermal_brane}.
In ref.\cite{thermal_brane}, it is discussed that an open string model can be
thermalized to stabilize a brane on the top of the different brane.
Unlike the usual models for brane inflation that starts with a large
expectation value for the inflaton field, thermal brane inflation starts 
at the top of an another brane.

\underline{Thermal brane inflation (original)}

Let us briefly review the idea of thermal brane
inflation proposed by Dvali\cite{thermal_brane}.
The following conditions are required so that the mechanism functions.

1) \,Exchange of the bulk modes such as graviton, dilaton or RR
fields govern the brane interaction at the large distance.

2)\, In the case when branes initially come close, bulk modes are in
equilibrium and their contribution to the free energy
can create a positive $T^{2}$ mass term for $\psi$ to stabilize the
branes on top of each other until the Universe cools down to a certain
critical temperature $T_{c}\sim m_{s}$.\footnote{
The author of ref.\cite{thermal_brane} 
considered open string modes stretched between different
branes.
When branes are on top of each other, these string modes 
are in equilibrium and their contribution to the free energy creates a
positive $T^2$ mass term. 
One may also worry about the problem of the domain wall formation.
In this model, domain wall formation is possible only when the broken
symmetry allows cosmological domain walls.
For example, domain walls may be formed when the number of
the transverse dimension is one, and at the same time the distant brane
B is not located at the boundary.
In such a case, the moving brane will have degenerated $\pm$ directions.
The above situation is already excluded in the original
model\cite{thermal_brane}. }
Here $m_s$ represents the negative curvature of $\psi$ at the origin,
 which is determined by the supersymmetry breaking.

The resultant scenario of thermal inflation is straightforward.
Assuming that there was a period of an early inflation with a reheat
temperature $T_{R}\sim M$, and at the end of inflation some of the
repelling branes sit on top of each other stabilized by the thermal
effects, one can obtain the number of e-foldings 
\begin{equation}
N_e=ln(\frac{T_{R}}{T_{c}}).
\end{equation}
Taking $T_{R}\sim 10 TeV$ and $T_{c}\sim 10^{3}- 10$ MeV, one finds 
$N_{e}\sim 10-15$, which is consistent with the original thermal
inflation\cite{thermal} and is enough to get rid of unwanted 
relics.\footnote{See also the previous arguments in ref.\cite{prev}}
In the original model the crucial restriction appears in $m_s$, which
must be large enough to satisfy the lower limit for the reheating
temperature.\footnote{For example, the reheating temperature for the
original thermal brane inflation is\cite{thermal_brane}
\begin{equation}
T_R \sim \sqrt{\frac{m_{osc}^3}{\phi_0^2}M_p}
\end{equation}
where $m_{osc}$ and $\phi_0$ are the mass of the oscillating inflaton
field and the vacuum expectation value of the inflaton at the true
vacuum, respectively.
Here we have omitted the numerical factors.
Of course, the negative mass at the origin ($m_s$) and the mass of the
oscillating inflaton ($m_{osc}$) is different.
However, as is discussed in the footnote 7, it seems hard to believe
that $m_s << m_{osc}$ is realized in natural settings.
Thus we assume $m_s > m_{osc}$.
In any case, the reheating temperature must be larger than the
temperature of the nucleosynthesis.
On the other hand, if one requires large number of e-foldings,
one needs small $m_s$ and large $\phi_0$ so that the thermal inflaton
lasts long. 
In the original model for thermal brane inflaton, 
$m_{osc} \sim 10 MeV >>m_{3/2}$
is assumed.
}
In our model, the above restriction is actually removed.

\underline{Hybrid alternative (Thermal hybrid inflation in the 
brane universe)}

Now it seems straightforward to improve the initial condition for
non-tachyonic brane inflation 
to fit the settings of thermal brane inflation.
To explain the idea, let us consider two branes (brane A and brane B)
fixed on some 
point in the extra dimensions, and a moving brane that is not fixed yet.
In the true vacuum, the moving brane stays on top of brane A, while 
the thermal effects confine it on top of brane
B during thermal inflation.
As we have discussed above, the initial condition for the non-tachyonic
brane inflation is satisfied within the above settings.
What we want to consider in this paper is a hybridization of thermal
brane inflation and non-tachyonic brane inflation.
Thermal hybrid inflation occurs if a moving brane, which is responsible
for the supersymmetry on brane A attaches to an another brane at a
distance.
If the components on each brane are appropriate to meet the
requirement from non-tachyonic brane inflation, thermal inflation starts.
At the end of thermal inflation, the moving brane falls apart from 
brane B and moves toward brane A.
When the moving brane come near to brane A, the spontaneously broken
supersymmetry is recovered and the inflation ends with the oscillation
of the field on the brane.

There are two unique characteristic features in this model.
In conventional models for thermal inflation, there is a lower limit for
the mass of the inflaton field that is derived from the requirement for
the reheating temperature.
This constraint is removed in our model because of the hybrid potential.
The second is that thermal inflation can become ``strong'' in our model.
To be more precise, we show why the conventional models for thermal
inflation were ``weak''.
In generic models for thermal inflation, inflation starts at 
$T_{in}\simeq \sqrt{m_{I}M}$, where $M$ and $m_I$ are the vacuum explain
value of the inflation field in the true vacuum and the mass of the
inflaton field, respectively.
Thermal inflaton ends at the temperature $T_{end}\simeq m_{I}$.
During this period, the Universe expands with the number of e-foldings of
\begin{equation}
N_e \simeq ln \left( \frac{T_{in}}{T_{end}} \right) \simeq
ln \left(\sqrt{\frac{M}{m_{I}}}\right).
\end{equation}
In this case, even if $M$ is as large as the GUT scale,
the number of e-foldings is at most $N_e\simeq 17$ for $m_I\simeq 1$GeV.
Thus we should conclude that the conventional model for thermal
inflation is a model for weak inflation, which cannot be used for the
first inflation. 

In our model, however, the situation is changed.
The energy density during inflation, which we denote by $V_0$, is
not related to the mass of the inflaton field.
The expected number of e-foldings is 
\begin{equation}
N_e \simeq ln \left(\frac{T_{in}}{T_{end}}\right) = 
ln\left(\frac{V_0^{1/4}}{m_{I}}\right).
\end{equation}
Although F-term inflation is successful in our model, one must
consider D-term inflation if one wants to protect inflaton mass from
corrections of O(H), where $H$ denotes the Hubble parameter.
For example, in models with $M_*>M_{GUT}$ and the gauge-mediated 
supersymmetry breaking 
on the brane, the number of e-foldings becomes
$N_e \simeq 42$ for $V_0^{1/4} \simeq 10^{12}GeV$ and $m_{I} \simeq 1 keV$.
On the other hand, when the fundamental scale is low,
one can obtain the number of e-foldings $N_e \simeq 35$ for 
$V_0^{1/4}\simeq 10^{6}$GeV and $m_I \simeq 10^{-2}$keV.\footnote{
Here we should explain how to get soft masses for the conventional
supersymmetric standard model.
In the above examples, we have assumed that the standard model fields 
and the source of the conventional supersymmetry breaking are 
localized on brane A.
We have also assumed that the supersymmetry breaking is mediated to the
standard model by
renormalizable interactions on brane A.
On the other hand, the source of the supersymmetry breaking during 
inflation can be located either on brane A or on the moving brane.
Of course the scale of the supersymmetry breaking during inflation is 
 higher and independent from the conventional (supersymmetric
standard model) supersymmetry breaking.
Considering the above conditions, we have assumed that the mass of
the inflaton near brane B during inflation is mediated by the
gravitational interactions.
All the sources of the supersymmetry breaking are assumed to be located 
on brane A.
If the source of the supersymmetry breaking is located on the moving
brane, the inflation mass ($m_I$) during thermal inflation will become
larger because during inflation the moving brane attaches to brane B, 
where excitation of light modes are 
expected\cite{thermal_brane}. 

The effective mass of the inflaton will be 
$m_I\simeq m_{3/2}<< \sqrt{V_0/M_p}$, if thermal brane
inflation is induced by the D-term.

In any case, the inflation mass ($m_I$) and the number of e-foldings
 ($N_e$) are strongly model
dependent.
We stress that $m_I$ is not the mass in the true vacuum, but the
 effective mass at a distance.} 

\section{Conclusions and Discussions}
\hspace*{\parindent}
In this paper we have constructed an example for thermal hybrid inflation.
In conventional thermal inflation, the requirement from the reheating
temperature puts a lower bound on the mass of the inflaton field.
The mass of the inflaton and the height of the potential during
inflation is intimately related to prevent ``strong'' inflation.
Our naive expectation is that one can remove the above constraints if 
 thermal inflation is realized within the setups for hybrid 
inflation.
With these things in mind, we have constructed a realistic model for
thermal hybrid inflation.
We have considered this problem in the scenarion of the brane world.
Our result drastically modifies the cosmological scenarios related to
thermal inflation.
For example, in ref.\cite{curvaton}, the curvaton hyposesis is 
discussed in the framework of thermal inflation.
In ref.\cite{curvaton}, they have concluded that the cosmological scales
cannot leave the horizon during ordinary thermal inflation, because the
constraint from the reheating temperature cannot meet the requirement.
However, as we have discussed above, such a constraint is removed
in our model.

We believe that our new model for thermal inflation opens up new
possibilities for the brane world.

\section{Acknowledgment}
We wish to thank K.Shima for encouragement, and our colleagues in
Tokyo University for their kind hospitality.


\begin{thebibliography}{1}
\bibitem{low_inflation}
N. Kaloper, A. Linde, Phys.Rev.D59:101303,1999;
N. Arkani-Hamed, S. Dimopoulos, N. Kaloper, and J. March-Russell,
Nucl.Phys.B567:189-228,2000;
R. N. Mohapatra, A. Perez-Lorenzana, and C. A. de S. Pires,
Phys.Rev.D62:105030,2000;
A. M. Green and A. Mazumdar, Phys.Rev.D65:105022,2002;
T. Matsuda, Phys.Rev.D66:107301,2002; Phys.Lett. B486 (2000) 300-305;
D. H. Lyth, Phys.Lett.B448:191-194,1999; Phys.Lett.B466:85-94,1999.
\bibitem{low_baryo}
G.R. Dvali, G. Gabadadze, Phys.Lett.B460:47-57,1999;
T. Matsuda, Phys.Rev.D65:103502,2002,
Phys.Rev.D66:023508,2002, Phys.Rev.D65:107302,2002,
Phys.Rev.D66:047301,2002; Phys.Rev.D64:083512,2001;
J.Phys.G27:L103-L108,2001; 
A. Masiero, M. Peloso, L. Sorbo, and R. Tabbash, Phys.Rev.D62:063515,2000;
A.Mazumdar, Nucl.Phys.B597(2001)561, Phys.Rev.D64(2001)027304;
A. Mazumdar and A. Perez-Lorenzana, Phys.Rev.D65:107301,2002; 
R. Allahverdi, K. Enqvist, A. Mazumdar, and A. Perez-Lorenzana,
Nucl.Phys.B618:277-300,2001;
A.Pilaftsis, Phys.Rev.D60:105023,1999;
R.Allahverdi, K.Enqvist, A.Mazumdar and A.P-Lorenzana,
Nucl.Phys. B618:377,2001;
S. Davidson, M. Losada, and A. Riotto, Phys.Rev.Lett.84:4284-4287,2000.
\bibitem{inverted}
D. H. Lyth, Ewan D. Stewart, Phys.Rev.D54:7186-7190,1996 
\bibitem{inverted_critical}
S.F. King and J. Sanderson, Phys.Lett.B412:19-27,1997 
\bibitem{matsuda_nontach2}
T.Matsuda, ``Non-tachyonic brane inflation'', hep-ph/0302035;
`` F-term, D-term and hybrid brane inflation'', hep-ph/0302078;
``Topological hybrid inflation in brane world'' hep-ph/0302204;
Phys.Rev.D65:103501,2002; Phys.Lett. B423 (1998) 35-39.
\bibitem{thermal_brane}
G.R. Dvali, Phys.Lett.B459:489-496,1999 
\bibitem{thermal}
D.H. Lyth and E. D. Stewart, Phys.Rev.D53:1784-1798,1996 
\bibitem{prev}
G. Lazarides, C. Panagiotakopoulos and Q. Shafi, Phys. rev. Lett.56,1986, 557;
G. Lazarides and Q. Shafi, Nucl. Phys. B392,1993,61
\bibitem{curvaton}
K. Dimopoulos and D. H. Lyth, 
``Models of inflation liberated by the curvaton hypothesis'', 
hep-ph/0209180 
\end{thebibliography}
\end{document}